\documentclass[11pt,superscriptaddress,amsmath,amssymb,nofootinbib,longbibliography]{revtex4-2}
\usepackage{graphicx}
\usepackage{dcolumn}
\usepackage{bm}
\usepackage{amssymb}
\usepackage{amsmath}
\usepackage{color}
\usepackage[normalem]{ulem}
\usepackage{dsfont}

\newcommand{\bwt}{\begin{widetext}}
\newcommand{\ewt}{\end{widetext}}
\newcommand{\newc}{\newcommand}
\newc{\hc}{\dagger}
\newc{\pd}{\partial}
\newc{\beq}{\begin{equation}}
\newc{\eeq}{\end{equation}}
\newc{\beqa}{\begin{eqnarray}}
\newc{\eeqa}{\end{eqnarray}}
\newc{\bi}{\begin{itemize}}
\newc{\ei}{\end{itemize}}
\newc{\ra}{\rightarrow}
\newc{\la}{\leftarrow}
\newc{\lra}{\longrightarrow}
\newc{\lla}{\longleftarrow}
\newc{\Lra}{\Longrightarrow}
\newc{\Lla}{\Longleftarrow}
\newc{\half}{\frac{1}{2}}
\newc{\fth}{\frac{1}{4}}
\newc{\hchecked}{\backslash\!\!\!\!\checkmark }
\newc{\del}{\delta}
\newc{\Del}{\Delta}
\newc{\gm}{\gamma}
\newc{\Gm}{\Gamma}
\newc{\lam}{\lambda}
\newc{\kap}{\kappa}
\newc{\tri}{\triangle}
\newc{\eps}{\epsilon}
\newc{\epsp}{\epsilon^\prime}
\newc{\ot}{\frac{1}{3}}
\newc{\tth}{\frac{2}{3}}
\newc{\ft}{\frac{4}{3}}
\newc{\wt}{\widetilde}
\newc{\ovl}{\overline}
\newc{\tchi}{\tilde{\chi}}
\newc{\ds}{\displaystyle}
\newc{\pmt}{\pm\!\pm}
\newc{\PL}{\hat{L}}
\newc{\PR}{\hat{R}}

\newc{\msm}{\mathrm{SM}}
\newc{\msh}{\mathrm{sh}}
\newc{\mtev}{\mathrm{TeV}}
\newc{\mgev}{\mathrm{GeV}}
\newc{\mmev}{\mathrm{MeV}}
\newc{\mkev}{\mathrm{keV}}
\newc{\mev}{\mathrm{eV}}
\newc{\Tr}{\mathrm{Tr}}
\newc{\nonr}{\nonumber}
\newc{\clbl}{\color{blue}}
\newc{\clg}{\color{green}}
\newc{\clr}{\color{red}}
\mathchardef\mhyphen="2D
\newc{\SL}{\not\!\!}
\usepackage{bm}
\usepackage{amsfonts}
\usepackage{xcolor}
\usepackage{amscd}
\usepackage{amsmath}

\begin{document}

\title{Non-universal gauged lepton number for charged lepton masses hierarchy and $(g-2)_{e,\mu}$}
\date{\today}

\author{We-Fu Chang}
\email{wfchang@phys.nthu.edu.tw}
\affiliation{Department of Physics, National Tsing Hua University, Hsinchu, Taiwan 30013, R.O.C. }

\begin{abstract}
We construct a novel flavor-dependent gauged lepton number $U(1)_\ell$ model for the hierarchical charged lepton masses and the observed $(g-2)_{e,\mu}$.
Only tau participates in the tree-level Standard Model ( SM ) Yukawa interaction.
At the same time, the masses of electron and muon are light due to radiative generation and(or) the heavy-mediator-suppressed Yukawa coupling to the SM Higgs. Not only can the measured anomalous magnetic dipole moment of the muon, $\tri a_\mu$, be explained, but the positive (or negative) $\tri a_e$ can also be accommodated  in this model.

Without additional discrete symmetries introduced,  charged lepton flavor violation is highly suppressed by the $U(1)_\ell$ symmetry.
Corresponding to two equally viable $U(1)_\ell$ charge assignments, this model predicts either  $A_{FB}^e >A_{FB}^\tau >A_{FB}^\mu$ or $A_{FB}^e <A_{FB}^\tau <A_{FB}^\mu$, which can be tested at the $e^+e^-$ machines before discovering the $U(1)_\ell$ gauge boson.
Moreover, the effective muon and electron Yukawa couplings can depart significantly from the SM predictions,  and those deviations could be probed at future $e^+e^-$ colliders and High-Luminosity LHC.

\end{abstract}
\maketitle

\section{Introduction}
The observed hierarchy among the charged fermion masses is one of SM's great puzzles.
In SM, the charged fermion masses stem from the SM Higgs Yukawa interaction with the predicted relationship $m_f= y_f v_0/\sqrt{2}$, where $y_f$ is the Yukawa coupling of fermion $f$ and $v_0\simeq 246\mgev$ is the vacuum expectation value(VEV) of the SM Higgs.
Experimentally, the Higgs Yukawa couplings of top\cite{ATLAS:2018hxb}, bottom\cite{ATLAS:2018kot, CMS:2018nsn}, tau\cite{ATLAS:2015xst,CMS:2017zyp}, and muon\cite{ATLAS:2020fzp,CMS:2020xwi}  are consistent,  within the errors of $\sim$ a few $\times 10\%$ \cite{ATLAS:2019nkf, CMS:2018uag, Zyla:2020zbs}, with the SM predictions.
The improvement of $y_\mu$  determination is promising at High-Luminosity LHC and future $e^+e^-$ colliders\cite{deBlas:2019rxi}, and an ultimate precision of $\sim 0.4\%$  is projected at the future FCC\cite{FCC:2018byv}.
Due to the smallness of $y_e$, six orders of magnitude smaller than $y_t$, the current upper limit on electron-Yukawa, $|y_e /y_e^{SM}|<260$ \cite{CMS:2014dqm,ATLAS:2019old}, is relatively poor. However, it is possible to reach a sensitivity of $|y_e /y_e^{SM}|<1.6$ at FCC-ee in the future\cite{dEnterria:2021xij}.

One interesting speculation about the charged fermion mass hierarchy is that the light charged fermion masses are radiatively generated.
Various models of this sort had been proposed a long time ago; see the review\cite{Babu:1989fg} and the earlier references therein. For more recent considerations along this line, see \cite{Dobrescu:2008sz, Okada:2013iba, Ma:2013mga, Gabrielli:2016vbb, Weinberg:2020zba, Ma:2020wjc, Baker:2021yli, Chiang:2021pma, Yin:2021yqy, Hernandez:2021iss, Adhikari:2021yvx, Jana:2021tlx, Mohanta:2022seo}.
For simplicity, we will only consider  the lepton sector and construct a model to justify the observed charged lepton mass hierarchy.

On the other hand, combining data from BNL E821 and the result of FNAL gives \cite{Abi:2021gix}
 \beq
 \tri a_\mu^{B\!N\!L-F\!N\!A\!L} = \tri a_\mu^{exp}-\tri a_\mu^{SM}\simeq (25.1\pm 5.9)\times 10^{-10}\,.
 \eeq
For a comprehensive review of the SM prediction of $(g-2)_\mu$, see \cite{Aoyama:2020ynm} and references therein\footnote{The SM prediction of the hadronic vacuum polarisation contribution to the muon $(g-2)$ is still under debate. The recent lattice calculations\cite{Borsanyi:2020mff,Ce:2022kxy,Alexandrou:2022amy, FermilabLattice:2022izv} predict a value consistent with the experimental result of muon $(g-2)$ but are in tension with the leading-order determination obtained by using dispersion relations, see \cite{Aoyama:2020ynm}.}.
As for the electron, $\tri a_e$ can be deduced with the input of fine-structure constant $\alpha_{em}$.
By adopting  $\alpha_{em}$ determined by using Cesium atoms\cite{Parker:2018vye} or Rubidium atoms\cite{Morel:2020dww}, $\tri a_e$  takes the value
\beq
\tri a_e^{Cs}  \simeq (-8.7\pm 3.6)\times 10^{-13}
\label{eq:deltaAe_minus}
\eeq
or
\beq
\tri a_e^{Rb}  \simeq (+4.8\pm 3.0)\times 10^{-13}\,,
\label{eq:deltaAe_plus}
\eeq
respectively.
Note the two values differ by $5.4\sigma$\cite{Morel:2020dww}, and more investigations are needed to settle the issue.
It is still unclear how to resolve this discrepancy in $\alpha_{em}$ determination at this moment, and thus
we will consider both, but separately, $\tri a_e^{Cs}$ and $\tri a_e^{Rb}$.

The radiative lepton mass generation mechanism closely connects to leptons' anomalous magnetic dipole moments.
As long as the Feynman diagrams for radiactive mass generation have charged degrees of freedom running in the loop, one can always attach an external photon to one of those charged particles and yield nonzero $\tri a_l$. Reversely, by removing the external photon line from the Feynman diagram(s) for $(g-2)$, a nonzero $\tri a_l$ implies  a finite  radiative mass correction to the lepton. The fact that $m_e, m_\mu \ll m_\tau$ and the observed deviations between $(g-2)_{\mu, e}$ and the SM predictions logically insinuate the possibility that electron and muon masses are radiatively generated.

We point out a novel anomaly-free gauged lepton $U(1)_\ell$ symmetry\footnote{See \cite{ Chao:2010mp, Schwaller:2013hqa, Chang:2018vdd,Chang:2018wsw, Chang:2018nid} for the early discussion on gauged lepton number.} to
rationalize the observed mass hierarchy $m_e,m_\mu \ll m_\tau$ and  accommodate the observed $\tri a_\mu^{B\!N\!L-F\!N\!A\!L}$ and $\tri a_e^{Cs[Rb]}$.
For recent similar  attempts, see, for example, \cite{Chiang:2021pma, Hernandez:2021iss}.
 The SM leptons carry different $U(1)_\ell$ charges in our model. Within the simple anomaly-free $U(1)_\ell$ charge assignment,  one
left-handed and one right-handed lepton are $U(1)_\ell$ neutral.
We identify the $U(1)_\ell$ neutral lepton as tau, which acquires its mass from the SM Yukawa coupling, while the SM electron and muon Higgs Yukawa interactions are forbidden.
 In addition to the exotic scalar sector, only one pair of exotic vector fermions, $N_{L,R}$, is introduced in this model.
The Dirac mass of vector fermion  $N_{L,R}$ plays a vital role in chirality flipping in generating  radiative masses and nonzero $\tri a$'s to electron and muon.

A nice feature of our model is the flexibility to separately fit $\tri a_\mu^{B\!N\!L-F\!N\!A\!L}$ with $\tri a_e^{Cs}$ or $\tri a_\mu^{B\!N\!L-F\!N\!A\!L}$ plus $\tri a_e^{Cs}$. This adaptability is due to the additional gauge-invariant electron Yukawa coupling to one of the exotic Higgs doublets.
Moreover,  ad hoc parities or symmetries are usually used to suppress the dangerous charged lepton flavor violating(CLFV) processes\footnote{ For a bottom-up approach to deal with the CLFV constrain, see \cite{Chang:2022eft}. } such as $l_i \ra l_j \gamma$, $l_i\ra l_a l_b l_c$, and $\mu\mhyphen e$ conversion.
In our model, the CLFV processes are highly suppressed or forbidden by the automatically emerged accidental symmetries.
For the flavor conserving part, this model predicts  that the muon and electron effective Yukawa couplings can deviate significantly from their SM predictions,
and such abnormal Yukawa couplings can be tested at the future colliders.

In Sec.\ref{sec:model}, the model is described in detail, and the radiative electron and muon mass generation and $(g-2)_{e,\mu}$ are also analyzed.
We demonstrate in Sec.\ref{sec:CLFV} that CLFV processes are highly suppressed and completely vanishing for tau flavor in this model.
In Sec.\ref{sec:pheno}, we discuss the phenomenology of the gauge boson and the effective electron- and muon-Higgs Yukawa couplings.
One of the important predictions is that the forward-backward asymmetries of $e,\mu,\tau$ are different and can be verified experimentally.
Before the conclusion, in Sec.\ref{sec:DM_Mnu}, we briefly address the possible extension of this model to include a dark matter candidate and the neutrino mass generation.

\section{Model}
\label{sec:model}
In this model, the SM leptons carry different $U(1)_l$ charges, see Table \ref{table:SMparticle}.
\begin{table}[htb]
\begin{center}
\begin{tabular}{|c||cccccc|c|}\hline
       &   \multicolumn{6}{c|}{SM lepton }  &  \multicolumn{1}{c|}{SM Higgs}\\\hline
 Symmetry $\backslash$ Fields & $L_\tau$ & $L_\mu$ & $L_e$ & $\tau_R$ & $e_R$ & $\mu_R$ & $H=\begin{pmatrix} H^+ \\ H^0\end{pmatrix}$   \\\hline\hline
$SU(2)_L$ &  \multicolumn{3}{c}{ $2$ }  & \multicolumn{3}{c|}{ $1$ } & $2$\\
$U(1)_Y$  & \multicolumn{3}{c}{$-\half$ } & \multicolumn{3}{c|}{ $-1$ } & $\half$\\  \hline
$U(1)_\ell$ & $0$  & $4$ & $-4$ &$0$ & $-7$ & $7$ & $0$ \\
\hline
\end{tabular}
\caption{The SM fields and their quantum numbers under the SM $SU(2)_L \otimes U(1)_Y $,  and the gauged lepton number $U(1)_\ell$. }
\label{table:SMparticle}
\end{center}
\end{table}
This model also calls for  one pair of vector fermion $N$, two doublets scalars, $H_{5,3}$, and three singlet scalars, $C_{8,6},S_3$, see Table \ref{table:newparticle}.
\begin{table}[htb]
\begin{center}
\begin{tabular}{|c||c|ccccc|}\hline
       &   \multicolumn{1}{c|}{ New Fermion}  &  \multicolumn{5}{c|}{ New Scalar}\\\hline
 Symmetry$\backslash$ Fields & $N_{L,R}$
  & $H_3=\begin{pmatrix} H_3^0 \\ H_3^-\end{pmatrix} $ & $H_5=\begin{pmatrix} H_5^0 \\ H_5^-\end{pmatrix} $ & $C_6$ & $C_8$ & $S_3$   \\\hline \hline
$SU(2)_L$ &  $1$& $2$ & $2$  & $1$  & $1$ & $1$\\
$U(1)_Y$  & $0$ &  $-\half$  & $-\half$ & $1$  & $1$ & $0$\\
$U(1)_\ell$ & $-1$ &  $-3$ & $5$ &$6$ & $-8$ & $3$ \\
\hline
\end{tabular}
\caption{New field content and quantum number assignment under the SM gauge symmetries $SU(2)_L \otimes U(1)_Y $,  and the gauged lepton numbers $U(1)_\ell$. }
\label{table:newparticle}
\end{center}
\end{table}
It is evident that our model is free of anomalies.  Note that it is still a viable solution if all $U(1)_l$ charges change sign. However, the sign is physical and can be determined by the interferences between the $U(1)_\ell$ gauge boson and the SM gauge bosons at the future $e^+e^-$ colliders.
The vector fermions admit a tree-level Dirac mass term $-M_N(\bar{N}_R N_L +\bar{N}_L N_R)$.

The most general gauge-invariant Yukawa interactions are
\beqa
{\cal L} &\supset&  y_{\mu R}\, \overline{N_L} \mu_R C_8 +  y_{\mu L}\, \overline{N_R}  H_5^\dagger  L_\mu +  y_{e R}\, \overline{N_L} e_R C_6 +
 y_{e L}\, \overline{N_R} H_3^\dagger L_e  \nonr\\
 &-&   y_\tau\, \overline{L}_\tau \tau_R H  + y_3\, \overline{L}_e e_R \wt{H}_3 +H.c.
\label{eq:L_Yukawa}
\eeqa
All the six new Yukawa couplings are complex in general, but only one phase combination is physical after field redefinition.
Note that the traditional global lepton number, where all $L_i$, $e_{Ri}$, and $N_{L,R}$ carry one unit of the conventional lepton number, is conserved.
In addition, this model acquires an accidental discrete symmetry $Z_\tau$ under which $L_\tau$ and $\tau_R$ are odd while all other fields transform trivially.  We emphasize that these two accidental symmetries, see Table \ref{table:acc_sym}, emerge automatically from the anomaly-free $U(1)_\ell$ charge assignment.
\begin{table}[htb]
\begin{center}
\begin{tabular}{|c||cc|c|c|}\hline
Accidental Symmetry$\backslash$ Fields & $L_\tau$ & $\tau_R$ & $L_\mu, L_e, \mu_R, e_R, N_{L,R}$
  & $H, H_3, H_5, C_8, C_6, S_3$   \\\hline \hline
global lepton number  & $1$ & $1$ &$1$ & $0$ \\
$Z_\tau$              &$-$  & $-$ &$+$  & $+$ \\
\hline
\end{tabular}
\caption{The fields and their quantum numbers under the accidental global lepton number and tau parity.}
\label{table:acc_sym}
\end{center}
\end{table}
The $U(1)_\ell$ is assumed to be spontaneously broken by $S_3$ at an energy scale higher than SM electroweak scale.
As $S_3$ takes its VEV, $\left\langle S_3 \right\rangle =v_3/\sqrt{2}$, the $U(1)_\ell$ gauge boson $X$ acquires a mass $M_X = 3 g_l v_3$, where $g_l$ is the unknown gauge coupling strength of $U(1)_\ell$.

Here, we focus on the relevant scalar mixing terms
\beq
{\cal L} \supset \mu_{H3} H^\dag \wt{H}_3 S^*_3  + \kappa_3 H^\dag H_3 S_3^* C_6 +  \kappa_5 H^\dag H_5 S_3 C_8  +H.c.\,,
\eeq
while the complete scalar sector lagrangian and some details can be found in Appendix \ref{app:scalarSEC}.
After the SSB of SM electroweak\footnote{We assume that $H_5$ and $H_3$ do not develop VEV.}, $\left\langle H_0 \right\rangle =v_0/\sqrt{2}$,  $H_5^+$ mixes with $C_8^+$, and $H_3^+$ mixes with $C_6^+$.
The two charged scalar sectors can be separately diagonalized by two rotations
\beq
U^{(l)}= \left(
           \begin{array}{cc}
            \cos \alpha^{(l)} & \sin \alpha^{(l)} \\
            -\sin \alpha^{(l)} & \cos \alpha^{(l)}  \\
           \end{array}
         \right)\,.
\eeq
The angles are given by
\beq
\sin 2\alpha^{(\mu)}= { \kappa_5 v_0 v_3 \over M_{H_{(e)}}^2-M_{L_{(e)}}^2 }\,,\;
\sin 2\alpha^{(e)}={ \kappa_3 v_0 v_3 \over M_{H_{(\mu)}}^2-M_{L_{(\mu)}}^2 }\,,
\eeq
where $M_{H_{(e/\mu)}}$ and $M_{L_{(e/\mu)}}$ are the mass eigenvalues of the heavier and lighter charged scalar associated with electron/muon, respectively.
Then in terms of the heavy(light) mass eigenstate $\Phi^{(e)}_{H(L)}$,
\beq
H_5^+ = \cos \alpha^{(\mu)}  \Phi^{(\mu)}_L + \sin \alpha^{(\mu)} \Phi^{(\mu)}_H\,,\;
C_8^+ = -\sin \alpha^{(\mu)} \Phi^{(\mu)}_L + \cos \alpha^{(\mu)} \Phi^{(\mu)}_H\,.
\eeq
Similarly, for the $H_3^+ \mhyphen C_6^+$ pair, we have
\beq
H_3^+ = \cos \alpha^{(e)}  \Phi^{(e)}_L + \sin \alpha^{(e)} \Phi^{(e)}_H\,,\;
C_6^+ = -\sin \alpha^{(e)} \Phi^{(e)}_L + \cos \alpha^{(e)} \Phi^{(e)}_H\,.
\eeq

\begin{figure}
\centering
\includegraphics[width=0.7\textwidth]{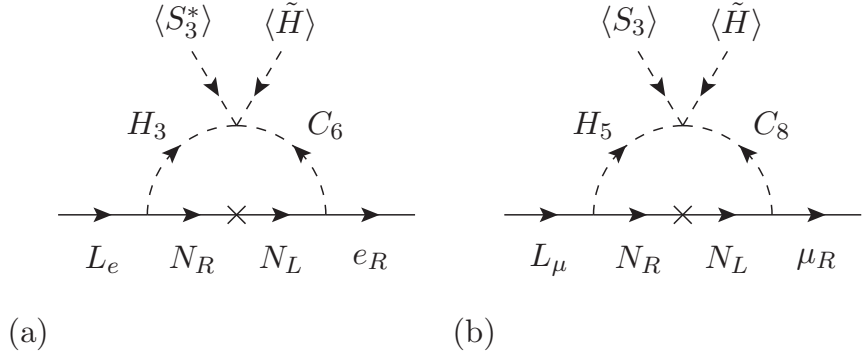}
\caption{ The Feynman diagrams ( in the interaction basis ) for $(g-2)_{e,\mu}$, effective Higgs couplings, and the radiative masses for electron and muon.
For $g-2$, the photon attaches to the charged scalar in the loop. For the radiative lepton-Higgs Yukawa couplings, one can either replace the Higgs VEV with  an external SM Higgs or connect the SM Higgs to the charged scalars.}
\label{fig:g-2_mass}
\end{figure}

The two 1-loop diagrams shown in Fig.\ref{fig:g-2_mass} give rise to electron and muon radiative masses. The corresponding radiative masses can be easily calculated as
\beqa
m_{loop}^{(e)} &=& { y_{e R}^*\, y_{e L} \over 32 \pi^2}\frac{\kappa_3 v_3 v_0}{M_N} {\cal F}_0\left(\beta^{(e)}_H, \beta^{(e)}_L \right)\,,\\
m_{loop}^{(\mu)} &=&{ y_{\mu R}^*\, y_{\mu L} \over 32 \pi^2}\frac{\kappa_5 v_3 v_0}{M_N} {\cal F}_0\left(\beta^{(\mu)}_H, \beta^{(\mu)}_L \right)\,,
\eeqa
where $\beta^i_{H/L}=\left(\frac{M^i_{H/L}}{M_N}\right)^2$, $i=(e), (\mu)$, and the definition of the loop function can be found in Appendix \ref{sec:loop_fun}. Note that  ${\cal F}_0(z_1, z_2)={\cal F}_0(z_2, z_1)>0$ if  $z_{1,2}\geq 0$.

\begin{figure}
\centering
\includegraphics[width=0.5\textwidth]{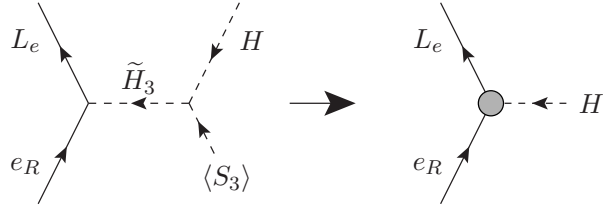}
\caption{ The Feynman diagram for generating the effective $e_R \mhyphen L_e \mhyphen H$ coupling after the SSB of gauged lepton number.}
\label{fig:SSS}
\end{figure}

In addition to the radiatively generated masses, the tree-level Yukawa interaction also yields an effective mass to electron, see Fig.\ref{fig:SSS} where the neutral component $H_3^0$ is the mediator.
Assuming the mixings between $H_3^0$ and other fields are small, we have
\beq
m^{(e)}_{tree}= {y^{(e)}_{tree} v_0 \over \sqrt2}\,,\;\;\mbox{where}\;\; y^{(e)}_{tree}\simeq {y_3 \mu_{H3} v_3 \over \sqrt2 M_3^2}\,,
\label{eq:m_tree}
\eeq
after integrating out the heavy $H_3^0$.
And tau acquires its mass $m_\tau = \frac{y_\tau v_0}{\sqrt{2}}$ via the SM Yukawa interaction.
Hence, the charged lepton mass matrix,
\beq
{\cal L} \supset -(\overline{ e_L},\overline{\mu_L},\overline{\tau_L})  \left(
      \begin{array}{ccc}
        m_{loop}^{(e)}+ m^{(e)}_{tree} & 0 & 0 \\
        0 & m_{loop}^{(\mu)} & 0 \\
        0 & 0 & \frac{1}{\sqrt{2}}y_\tau v_0 \\
      \end{array}
    \right)
    \left(
      \begin{array}{c}
        e_R \\ \mu_R \\ \tau_R
      \end{array}
    \right) + H.c.\,,
\eeq
 is diagonal in this model at the 1-loop level. And it will be clear that the SM CLFV dim-4 operators, $\overline{L}_i e_{Rj} H\, (j\neq i)$, vanish to all orders.

For muon, the phase of $m_{loop}^{(\mu)}$ can be removed by muon chiral field redefinition.
Since ${\cal \wt{F}}_0(\beta^{(\mu)}_H, \beta^{(\mu)}_L )>0$, we can always choose the combination $ \kappa_5 (y_{\mu R}^* y_{\mu L} )$ as a positive real number such that the muon mass is positively defined.
Similarly, the phase of the complex $y_\tau$ can also be absorbed by  redefinition of the tau chiral fields so that $y_\tau$ takes a positive real value $ y_\tau=\sqrt{2} (m_\tau \mgev/v_0) \simeq 1.02\times 10^{-2}$.

With one photon attached to the charged scalar in the loop diagrams shown in Fig.\ref{fig:g-2_mass},  one can calculate the anomalous magnetic moments of electron and muon.
For $m_F\gg m_l$, we have\cite{Chang:2021axw}
\beqa
\tri a_e &=&  { \Re[ y_{e R}^*\, y_{e L} ]  \over 32 \pi^2} \frac{m_e}{M_N^2} \frac{\kappa_3 v_3 v_0}{M_N}
{\cal \wt{I}}_0\left(\beta^{(e)}_H, \beta^{(e)}_L \right)\,,\nonr\\
\tri a_\mu &=&  { \Re[ y_{\mu R}^*\, y_{\mu L} ]  \over 32 \pi^2} \frac{m_\mu}{M_N^2} \frac{\kappa_5 v_3 v_0}{M_N}
{\cal \wt{I}}_0\left(\beta^{(\mu)}_H, \beta^{(\mu)}_L \right)\,,
\eeqa
where $m_{e,\mu}$ are the physical lepton masses, and the loop function is given by
\beq
{\cal \wt{I}}_0 \equiv {{\cal J}_0(z_1) -{\cal J}_0(z_2) \over  z_1- z_2} \,,\; \mbox{and}\;\;
{\cal J}_0( z ) ={1-z^2+2z\ln z \over (z-1)^3 }\,.
\label{eq:a2J}
\eeq
Note ${\cal \wt{I}}_0$ is symmetric and positively defined.

\begin{figure}
\centering
\includegraphics[width=0.4\textwidth]{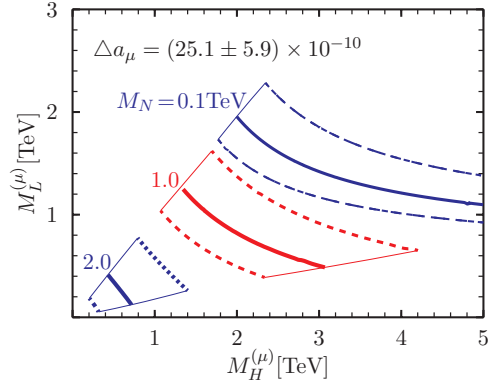}
\caption{ The 2-dimensional parameter space of $M_{H/L}$ for the muon.
The solid(dash) lines represent $\tri a_\mu$ at the central($1\sigma$ boundary) values for
$M_N=\{0.1, 1.0, 2.0\}\mtev$.}
\label{fig:a_mu}
\end{figure}

\begin{figure}
\centering
\includegraphics[width=0.4\textwidth]{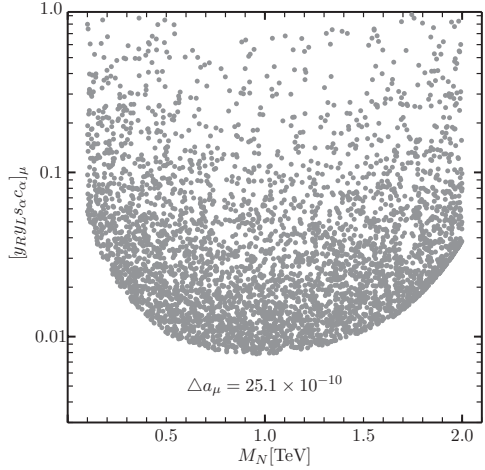}
\caption{ The mixing product $y_{\mu R}^* y_{\mu L}\sin \alpha^\mu \cos \alpha^\mu$ for muon v.s. $M_N$ for $\tri a_\mu =25.1 \times 10^{-10}$.}
\label{fig:mu_mix}
\end{figure}

For electron and muon, the ratio
\beq
\frac{\tri a_{l}}{m_{loop}^{(l)}} = \frac{m_{l}}{M_N^2}{ \Re[ y_{l R}^*\, y_{l L} ]  \over y_{l R}^*\, y_{l L} }  {{\cal \wt{I}}_0\left(\beta^{(l)}_H, \beta^{(l)}_L \right) \over  {\cal F}_0\left(\beta^{(l)}_H, \beta^{(l)}_L \right) }
\eeq
is independent of the mixing angle, or equivalently $\kappa_{5,3}$.
For muon, $y_{\mu R}^*\, y_{\mu L}$ can be made a real number by field redefinition or $m_{loop}^{(\mu)} =m_\mu \simeq 0.105\mgev$. Hence, $\tri a_\mu>0$ follows and agrees with the sign of the measured $\tri a_\mu^{B\!N\!L-F\!N\!A\!L}$.
On the other hand,  the tree-level Yukawa contribution mediated by $H_3$ is reserved for the electron, so either positive or negative $\tri a_e$( $m_{loop}^{(e)}$) could be accommodated.
Moreover, since $m_\mu$ is known, $\tri a_\mu$ only depends on three physical masses, $M_N$ and $M^{(\mu)}_{H/L}$.
In Fig.\ref{fig:a_mu} we display the allowed 2-dimensional parameter space of $M^{(\mu)}_{H/L}$, for a given $M_N$, which gives rise to the observed $1\sigma$ range of $\tri a_\mu^{B\!N\!L-F\!N\!A\!L}$. As seen in Fig.\ref{fig:a_mu}, the typical values for $M^{(\mu)}_{H/L}$ are about a few $\mtev$.
We also show the product $y_{\mu R}^* y_{\mu L} s_\alpha^{(\mu)}c_\alpha^{(\mu)}$ vs $M_N$ in Fig.\ref{fig:mu_mix}. For a fixed $M_N$, we scan the region of $ 0.2\mtev < M_{L_{(\mu)}}< M_{H_{(\mu)}}<5\mtev$ to find the viable solution which yields the measured $\tri a_\mu^{B\!N\!L-F\!N\!A\!L}$.
The typical value of the mixing product is in the reasonable range of $10^{-2}- 10^{-1}$ for $M_N\in[0.1, 2.0]\mtev$, and the minimum, $\sim 10^{-2}$, happens at around $M_N\sim 1\mtev$. By assuming $|y_{\mu R,\mu L}|<\sqrt{4\pi}$, we obtain a lower bound $ |\sin \alpha^{(\mu)}|  \gtrsim 10^{-3}$ for $\tri a_\mu^{B\!N\!L-F\!N\!A\!L}$, which is to be accommodated in this model.

It is  evident that the new physics(NP) contribution to $\tri a^{NP}_\tau=0$ at the 1-loop level.
The leading contribution to $\tri a^{NP}_\tau$ is the 2-loop Barr-Zee-like diagrams with charged scalars running in the loop, which gives rise to the effective $\gamma\gamma H_{SM}$ vertex. In this model, the charged scalars are $\phi_{H/L}^{(e,\mu)}$.  The ballpark estimation gives
\beq
\tri a^{NP}_\tau \sim \frac{\alpha_{em}}{(4\pi^2)^2}\sum_k \frac{\mu_k}{v_0}\frac{m_\tau^2}{M_k^2} \ln \frac{M_h^2}{M_k^2}\,,
\eeq
where $\mu_k$ represents  the triple coupling of the charged scalar, of mass $M_k$, with the SM Higgs,  and $M_h$ the SM Higgs mass.
Taking $\mu_k \sim v_0$, $M_k\sim {\cal O}(\mtev)$,  this model predicts $|\tri a^{NP}_\tau| \lesssim {\cal O}(10^{-10})$, which is beyond the experimental sensitivities in the near future.

\begin{figure}
\centering
\includegraphics[width=0.4\textwidth]{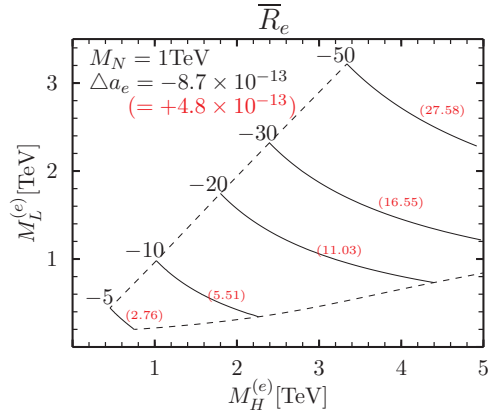}
\caption{ The contour (solid lines) plot of $\overline{R}_e=  m^{(e)}_{loop}/m_e$, while the dashed lines indicate the boundary.
We take the values $M_N=1$ TeV and $\tri a^{Cs}_e=-8.7\times 10^{-13}$ as the reference.
For different values of $\tri a_e$, the contour values scale linearly  in $\tri a_e$.
For example, the values in parentheses are for $\tri a^{Rb}_e= +4.8\times 10^{-13}$.}
\label{fig:a_e}
\end{figure}

Now we turn our attention to the electron. Since one can only remove one physical CP phase through the field redefinition, it has to be reserved for making the physical mass real positive.
Similar to the calculation of  $\tri a_e$, the electric dipole moment of the electron from NP can be derived  as
\beq
d^{NP}_e = - e \frac{ \Im[ y_{e R}^*\, y_{e L} ]}{64 \pi^2} {   \kappa_3 v_0 v_3 \over  M_N^3}
{\cal \wt{I}}_0\left(\beta^{(e)}_H, \beta^{(e)}_L \right)\,.
\eeq
In terms of $\tri a_e$, it can also be expressed as
\beq
d^{NP}_e = - {e \tan \delta^{(e)}_{CP} \over 2 m_e} \tri a_e\,,\; \mbox{where}\;\;\delta^{(e)}_{CP} ={\rm{arg}} (y_{e R}^*\, y_{e L})\,.
\eeq
Note that the $m_e$  here is the positive real physical mass.
Plugging in the value of $\tri a_e = -8.7[+4.8]\times 10^{-13}$ and the latest limit on $|d_e|<1.1\times 10^{-29} e\mhyphen cm$\cite{Andreev:2018ayy}, we obtain a stringent bound that
\beq
\left|\tan \delta^{(e)}_{CP} \right|<6.56[3.62] \times 10^{-7}\,.
\eeq
The phenomenological consideration indicates that the two phases of $m^{(e)}_{tree}$ and $m^{(e)}_{loop}$ are aligned to the level of ${\cal O}(10^{-7})$. The smallness of the relative phase, $\delta^{(e)}_{CP}$,  cannot be addressed in the current model setup. It suggests that CP symmetry should be assumed, at least for the lepton sector.
Therefore, in the rest of the paper, we set $\delta^{(e)}_{CP}=0$ for simplicity\footnote{
Note that, in this model, $d^{NP}_\mu=d^{NP}_\tau =0$ at the 1-loop level  even without assuming CP symmetry.
On the other hand, if both $\tri a_e$ and $\tri a_\mu$ are positive, one can exchange the identities of electron and muon such that $d^{NP}_e=d^{NP}_\tau =0$ at the 1-loop level. In that case, $d^{NP}_e$ starts at the 3-loop level and the constraint from $d^{NP}_e$ is much alleviated. Moreover,  $d^{NP}_\mu = 2.4\times 10^{-22} \times\tan \delta^{(\mu)}_{CP}\,\mbox{e-cm}$  is safely below the current bound $|d^{NP}_\mu|<1.8\times 10^{-19}\,\mbox{e-cm}$ \cite{Muong-2:2008ebm}.
}.
Then, the overall phase can be removed by electron filed redefinition so
\beq
m_e= y^{(e)}_{tree} { v_0 \over \sqrt2}  +\tri a_e \frac{M_N^2}{m_e} \wt{R}\left(\beta^{(e)}_H, \beta^{(e)}_L \right)  \,,\; \mbox{ where }\;\;
\wt{R}(z_1,z_2)\equiv {  {\cal F}_0\left(z_1, z_2\right) \over {\cal \wt{I}}\left(z_1,z_2 \right) }\,
\eeq
and $\wt{R}>0$.
Equivalently, we have
\beq
y_3 
 =  \frac{ 6 g_l}{\sqrt{2}} {M_3^2 \over \mu_{H3} M_X  } \left ( 1 - \overline{R}_e  \right ) y^{(e)}_{SM}\,,\;
  \overline{R}_e \equiv \frac{ m^{(e)}_{loop}}{m_e} = \tri a_e \frac{M_N^2}{m_e^2}\wt{R}\left(\beta^{(e)}_H, \beta^{(e)}_L \right)\,,
\eeq
where $ y^{(e)}_{SM}= \sqrt{2}m_e/v_0= 2.94\times 10^{-6}$ is the SM electron Yukawa coupling.
From Fig.\ref{fig:a_e}, we see the typical value of $| \overline{R}_e|$, the ratio of the radiative mass to the physical mass of electron, is about $ \sim {\cal O}(10)$ for $M_{H,L}^{(e)} \sim \mtev$.
If we take $M_N\simeq M_X \simeq 1 \mtev$, $M_3(\sim M^{(\mu)}_H ) \simeq 2 \mtev$, and $\overline{R}_e \simeq -10$, then
\beq
y^{Cs}_3
\sim 1.497\times 10^{-2}\times  \left( {g_l \over 0.1 e }\right) \times  \left( {10\mgev \over \mu_{H3}}\right)\,.
\eeq
On the other hand, if one adopts  $\tri a_e^{Rb}=4.8\times 10^{-13}$ and keeps all other parameters fixed, then
\beq
y^{Rb}_3 \sim -0.612 \times 10^{-2}\times  \left( {g_l \over 0.1 e }\right) \times \left( {10\mgev \over \mu_{H3}}\right)\,.
\eeq
Compared with $y^{(\tau)}_{SM}=1.02\times 10^{-2}$, this model does not need ridiculous fine tuning to yield the observed charged lepton mass hierarchy.

\section{Charged lepton flavor violation}
\label{sec:CLFV}
First, note that both the traditional lepton number and the accidental $Z_\tau$ parity are intact after the SSB of $U(1)_\ell$.
If we follow the fermion line of an incoming tau for any Feynman diagram, it must end up with an outgoing tau.
Therefore, there is no tau number violation in this model to all orders, and we only need to consider the CLFV in the electron and muon sectors.
Because the SM gauge symmetries are at work in the intermediate energy scale between the SSB of $U(1)_\ell$ and the SSB of the SM electroweak, we address the possible CLFV in the $e-\mu$ sector by considering the SM gauge-invariant operators in terms of SM DOF.

\begin{figure}[htb]
\centering
\includegraphics[width=0.95\textwidth]{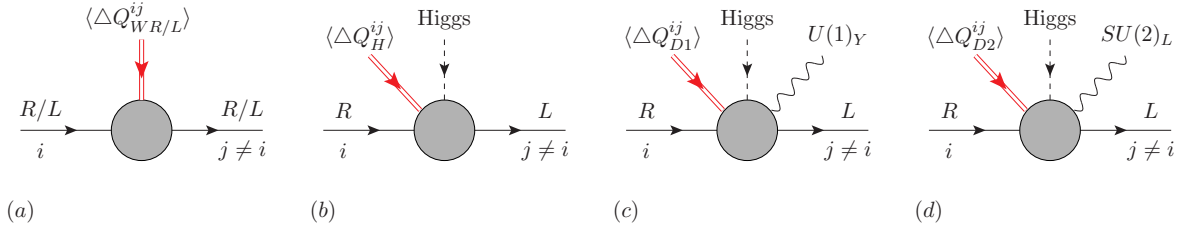}
\caption{ The dim-4 (a,b) and the dim-6 (c,d) dipole charged lepton flavor changing operators and the required quantum number injection (double line).
The gray blobs represent any possible perturbative Feynman diagrams that constitute the vertices. The indices $i,j$ stand for flavor and $L/R$ label the charged lepton chirality. }
\label{fig:FCNC_3}
\end{figure}

Starting from dim-4, we need to consider only three operators (and their hermitian conjugations),
\beq
\overline{L}_\mu \gamma^\alpha D_\alpha L_e\,,\;
\overline{\mu}_R \gamma^\alpha D_\alpha e_R\,,\;
\overline{L}_\mu e_R H\,,
\label{eq:dim4OP}
\eeq
where $D_\alpha$ is the SM covariant derivative. The first two operators give rise to the CLFV wavefunction corrections, while the last one generates the cross-flavor mass term below the SSB of SM electroweak.
The most general Feynman diagrams consist of two external leptons can be pictorially illustrated in Fig.\ref{fig:FCNC_3}(a,b), where
the gray blobs represent any possible perturbative Feynman diagrams that constitute the vertices.
The corresponding $U(1)_\ell$ charges for operators listed in Eq.(\ref{eq:dim4OP}) are $\{ -8, -14, -11\}$, respectively.
In this model, only $H$ and $S_3$ are higgsed, thus the injected $U(1)_\ell$ charge can only be $\tri Q_l=3 k$, where $k$ is an integer,
corresponding to the number of external $S_3$ lines above the SSB of $U(1)_\ell$.
Therefore, all the dim-4 CLFV operators are forbidden to all orders in this model. And the resulting $h_{SM} \ra \mu e$ from the dim-4 operators also vanishes.

\begin{figure}[htb]
\centering
\includegraphics[width=0.8\textwidth]{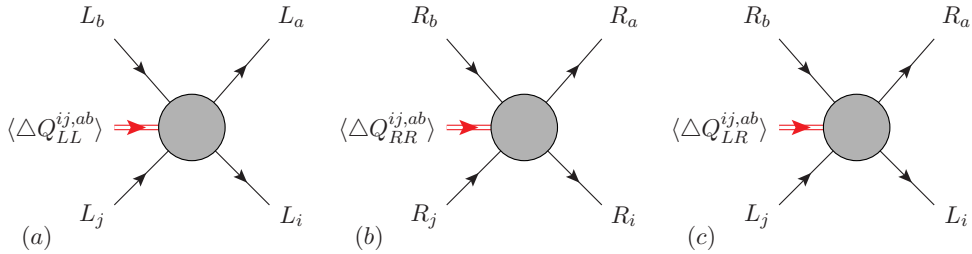}
\caption{ The general Feynman diagrams for the CLVF 4-lepton operators and the required quantum number injection(double line).
The gray blobs represent any possible perturbative Feynman diagrams that constitute the vertices.The indices $i, j, a, b$ stand for flavor, and $L/R$ label the charged lepton chirality. }
\label{fig:FCNC_4}
\end{figure}

We move on to consider the dim-6 operators.
The complete\footnote{  Note that $(\bar{L}_i \sigma^{\alpha \beta} R_j)(\bar{R}_a \sigma_{\alpha \beta}L_b)=0$, $\left(\overline{L}_i \gamma^\alpha \overrightarrow{\tau} L_j \right)\left(\overline{L}_a \gamma^\alpha \overrightarrow{\tau} L_b\right)= 2\wt{O}_{LL}^{ibaj}- \wt{O}_{LL}^{ijab}$, and $-2(\bar{L}_i R_j)(\bar{R}_a L_b) = \wt{O}_{LR}^{ibaj}$, by using the Fierz transformation and the properties of Pauli matrices.} SM gauge-invariant dimension-6 dipole, Fig.\ref{fig:FCNC_3}(c,d), and 4-lepton operators, Fig.\ref{fig:FCNC_4}, are
\beqa
&& \wt{O}_{D1}^{ij}= \left(\overline{L}_i \sigma^{\alpha \beta} R_j \right)H  F_{\alpha \beta} \,,\;
\wt{O}_{D2}^{ij}= \left(\overline{L}_i \sigma^{\alpha \beta}\overrightarrow{\tau} R_j\right) H  \overrightarrow{W}_{\alpha \beta}\,,\nonr\\
&& \wt{O}_{LL}^{ijab}=\left(\overline{L}_i \gamma^\alpha L_j \right)\left(\overline{L}_a \gamma^\alpha L_b\right)\,,\,
\wt{O}_{LR}^{ijab}=\left(\overline{L}_i \gamma^\alpha L_j \right)\left(\overline{R}_a \gamma^\alpha R_b\right)\,,\,
\wt{O}_{RR}^{ijab}=\left(\overline{R}_i \gamma^\alpha R_j \right)\left(\overline{R}_a \gamma^\alpha R_b\right)\,,
\eeqa
where $\vec{\tau}$ is the Pauli matrix, $F_{\alpha\beta}$ and $W_{\alpha \beta}$ are the field strengthes of $U(1)_Y$ and $SU(2)_L$, respectively.
The operators $\wt{O}_{D1,D2}^{\mu e}$ need $+11$  $U(1)_\ell$ charge injection and are forbidden to all orders.
As a result, the dimension-6 contributions to $\mu\ra e \gamma$ and $Z\ra \mu e$ vanish.

For the CLFV 4-lepton operators, the required quantum numbers injection ( double line in Fig.\ref{fig:FCNC_4} ) into these blobs are listed in Table \ref{table:CLFVEO}, where the $(e\mu)(ee)$ column relates to $\mu\ra 3e$, and the $(e \mu)(e \mu)$ column relates to muonium-antimuonium oscillations.
Again, only $H$ and $S_3$ are higgsed in this model, and none of the required quantum number injection is possible to all orders.
One can  easily generalize the analysis to the dim-6 2-lepton-2-quark operators and conclude there is no CLFV 4-fermion operator to all orders in this model setup. The dim-6 operator for $\mu\mhyphen e$ conversion on nuclei is also forbidden.
\begin{table}[htb]
\begin{center}
\begin{tabular}{|c|c|c|c|}
  \hline
  operator $\diagdown$ (ij)(ab) & $(e \mu)(l_k l_k)$  & $(e \mu)(e \mu)$  & $( \tau e)(\mu\tau)$  \\
  \hline
  $\wt{O}_{LL}^{ijab}$ & $8$ &$ 16$ &$-8$\\
  $\wt{O}_{RR}^{ijab}$ & $14$ &$28$ &$-14$\\
 $\wt{O}_{LR}^{ijab}$ & $8$ &$22$ &$-11$\\
 $\wt{O}_{RL}^{ijab}$ & $14$ &$22$ &$-11$\\
  \hline
\end{tabular}
\caption{The net $U(1)_\ell$ charge of the CLFV dimension-six 4-lepton operators, where $l_k=e,\mu,\tau$, and $i,j,a,b$ are the flavor indices.  }
\label{table:CLFVEO}
 \end{center}
 \end{table}

 In principle, the RGE evolution could induce
CLFV  below the electroweak scale from higher dimensional($>6$) operators\footnote{ Since only leptons are charged under $U(1)_\ell$, the next order of CLFV relevant operators are dim-10 6-lepton ones.
For example, the CLFV vertex with three incoming muons and three outgoing electrons plus one external SM Higgs, which leads to CLFV 2-to-4 scattering $e^-e^-\ra e^+\mu^-\mu^-\mu^-$,  is possible by proper $v_3$ insertion.  Although this general vertex is symmetry allowed, we failed to find the corresponding Feynman diagram(s). }.
Although the general discussion on CLFV operators with dimensions higher than six is beyond the scope of this paper, we expect
the RGE running effects to be suppressed and insignificant.

\section{Phenomenology}
\label{sec:pheno}
By convention,  we will take $g_2, g_1, g_l$, the gauge coupling strengths, to be positive.
Since the SM parts in the covariant derivative
\beq
D_\mu = \partial_\mu -i g_2 |T_3|\,\vec{\sigma}\cdot \vec{A}_\mu - i g_1 Y B_\mu -i g_l Q_l X_\mu
\eeq
stay the same, we focus on the new gauge $U(1)_l$ interaction part. The leptons interact with
the new gauge boson $X$ via
\beq
{\cal L} \supset -i g_l \sum_l\left[ \bar{l} \gamma_\mu( G_{lL}\PL+ G_{lR}\PR) l \right] X^\mu
\eeq
with $-G_{e L}= G_{\mu L}=4$, $-G_{e R}= G_{\mu R}=7$, and $G_{\tau L}= G_{\tau R}=0$.

If integrating out the heavy $X$, one obtains an effective contact interaction
$\frac{16 g_l^2}{M_X^2}(\bar{e}_L \gamma^\alpha e_L)(\bar{\mu}_L \gamma_\alpha \mu_L)$.
From the limit\cite{Zyla:2020zbs} $ \Lambda^+(ee\mu\mu)>8.5\mtev$, we get
\beq
g_l <\sqrt{\frac{\pi}{8}} \left( {M_X \over 8.5 \mtev}\right)\simeq 0.24 e \left( {M_X \over \mtev}\right)\,.
\label{eq:contactbound}
\eeq
The  widthes of $X$ decays into fermion and scalar pairs are given by
\beqa
\Gamma_{X\ra l^+l^-} &=& \Theta(1-4\beta_l) \frac{g_l^2}{24\pi} M_X [G_{lL}^2+G^2_{lR}+6\beta_f G_{lL} G_{lR}]\sqrt{1-4\beta_l}\,,\\
\Gamma_{X\ra SS^*}&=& \Theta(1-4\beta_S) \frac{g_l^2 Q_s^2}{48\pi} M_X (1-4\beta_S)^{\frac{3}{2}}\,,
\eeqa
respectively.
In the above, $Q_s$ is the $U(1)_\ell$ charge of the scalar $S$, $\beta_i = (m_i/M_X)^2$, and $\Theta$ is the step function.
If all the exotic scalars are heavier than $M_X/2$, the decay width of $X$ is dominated by the modes with 2-body $e^+e^-, \mu^+\mu^-,\bar{\nu}_e{\nu}_e, \bar{\nu}_\mu{\nu}_\mu $ final states. One has
\beq
{\Gamma_X \over M_X} \simeq 27 \alpha_{em} \left(\frac{ g_l}{e} \right)^2 < 1.17\times 10^{-2} \left( {M_X \over \mtev}\right)^2\,,
\eeq
where $\alpha_{em}$ is the fine structure constant, and the upper bound stems from Eq.(\ref{eq:contactbound}).
At the $X$-pole, the narrow-width gauge boson has a large cross-section
\beq
\sigma(e^+e^-\ra X^*\ra l^+ l^-) ={g_l^4 (G^2_{eL}+G^2_{eR} )^2\over 24\pi M_X^2}\frac{M_X^2}{\Gamma_X^2}\,,
\eeq
where $l=e, \mu$.
For $M_X=1\mtev$ and $g_l=0.1 e$, this cross-section is about $6.1\times 10^{10} \,fb$.
If $X$ boson can be produced at the future high energy $e^+e^-$ collider,
the narrow peaks of invariant masses $m_{ee} \sim m_{\mu\mu}\sim M_X$ will be smoking gun evidence of the gauge $U(1)_\ell$.
On the other hand, even below the $X$ resonance, the interferences between $X$ and the SM gauge boson cause significant differences in
the forward-backward asymmetry of leptons. The formulae of forward-backward asymmetry of leptons have been collected in Appendix \ref{sec:appendAFB}. In Fig.\ref{fig:AFB}, we display the $A_{FB}$ of three leptons by assuming that $M_X=1\mtev$, $g_l=0.1e$, $Q_l^N=-1$, and $X$ only decays into a lepton pair.
Since tauon does not couple to $X$, it follows the SM prediction. Due to the different $U(1)_l$ charges,  $A^e_{FB}$ and $A^\mu_{FB}$ differ from the SM prediction significantly.  At the Z-pole, $A^e_{FB}/A^\tau_{FB}\sim 1.0001$ and $A^\mu_{FB}/A^\tau_{FB}\sim 0.9999$. However, the current experimental precision cannot tell the differences at $Z$-pole.
The differences grow and reach $\sim{\cal O}(1)$ as $\sqrt{s}$ increases and approaches $M_X$.
This prediction is robust and can be tested by future $e^+e^-$ colliders.

\begin{figure}
\centering
\includegraphics[width=0.45\textwidth]{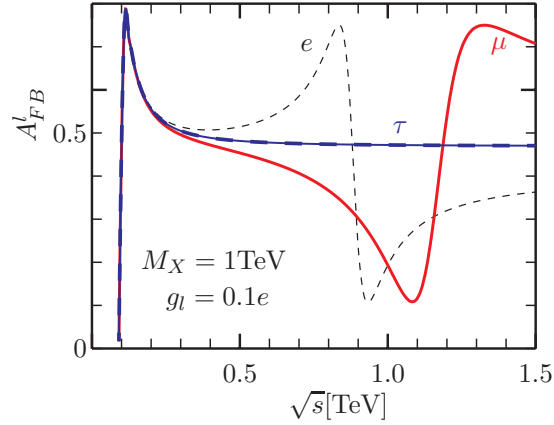}
\caption{Lepton forward-backward asymmetry for $M_X=1\mtev$, $g_l= 0.1 e$, and $Q_l^N=-1$.
If one adopts the other $U(1)_l$ charge assignment with the signs of all $U(1)_\ell$ charges flipped, then $A^e_{FB} \Leftrightarrow A^\mu_{FB}$. }
\label{fig:AFB}
\end{figure}

It is well-known that the oblique parameters\cite{Peskin:1990zt} $S$ and $T$ constrain the mass splitting of the doublet scalars.
In this model, $H_5$ and $H_3$ contribute
\beq
\tri T_i =\frac{1}{16\pi s_W^2}\frac{M_{i+}^2}{M_W^2}\left[ 1+z_i +{2z_i\over 1-z_i}\ln z_i\right]\,,\;
\tri S_i= -\frac{1}{12\pi}\ln z_i\,,\;\;(i=H_5, H_3)\,,
\eeq
where $z_i= (M_{i-}/M_{i+})^2$ is the ratio of $T_3=-1/2$ charged component mass squared to that of the $T_3=1/2$ neutral component.
Expanding around the degenerate case and denote $\tri M_i=M_{i+}-M_{i-}$,
\beq
\tri T_i \simeq \frac{1}{12\pi s_W^2}\frac{(\tri M_i)^2}{M_W^2}\,,\;
\tri S_i \simeq \frac{1}{6\pi} \frac{\tri M_i}{M_{i+}}\,.
\eeq
$\tri S_i$ can be either positive or negative, but $\tri T_i$ is always positive.
Thus, without any sign ambiguity, $\tri T= \tri T_5 +\tri T_3$ can be used to constrain the model parameters.
From $T_{exp}<0.22$ at $95\%$ C.L.\cite{Zyla:2020zbs},
\beq
\tri M_5^2+ \tri M_3^2 <1.91 M_W^2\,.
\eeq
Compared to $M_5, M_3\sim {\cal O}(\mtev)$, both $H_5$ and $H_3$ have small mass splitting, $\lesssim M_W$. This implies that
$\tilde{\lambda}_{H3/H5} \simeq \bar{\lambda}_{H3/H5}$\footnote{ Another possibility is  $ |\tilde{\lambda}_{H3}|, |\tilde{\lambda}_{H5}|, |\bar{\lambda}_{H3}|, |\bar{\lambda}_{H5}|\ll 1$. Due the presence of $\kappa_{5,3}$ terms, one has to  check numerically
that the scalar potential is bounded from below. } and
$|\kappa_{5,3}|< 1$ so that the mixings between $H_{5,3}$ and other scalars are small.

\subsection{Effective Yukawa couplings}
\begin{figure}[htb]
\centering
\includegraphics[height=0.22\textheight]{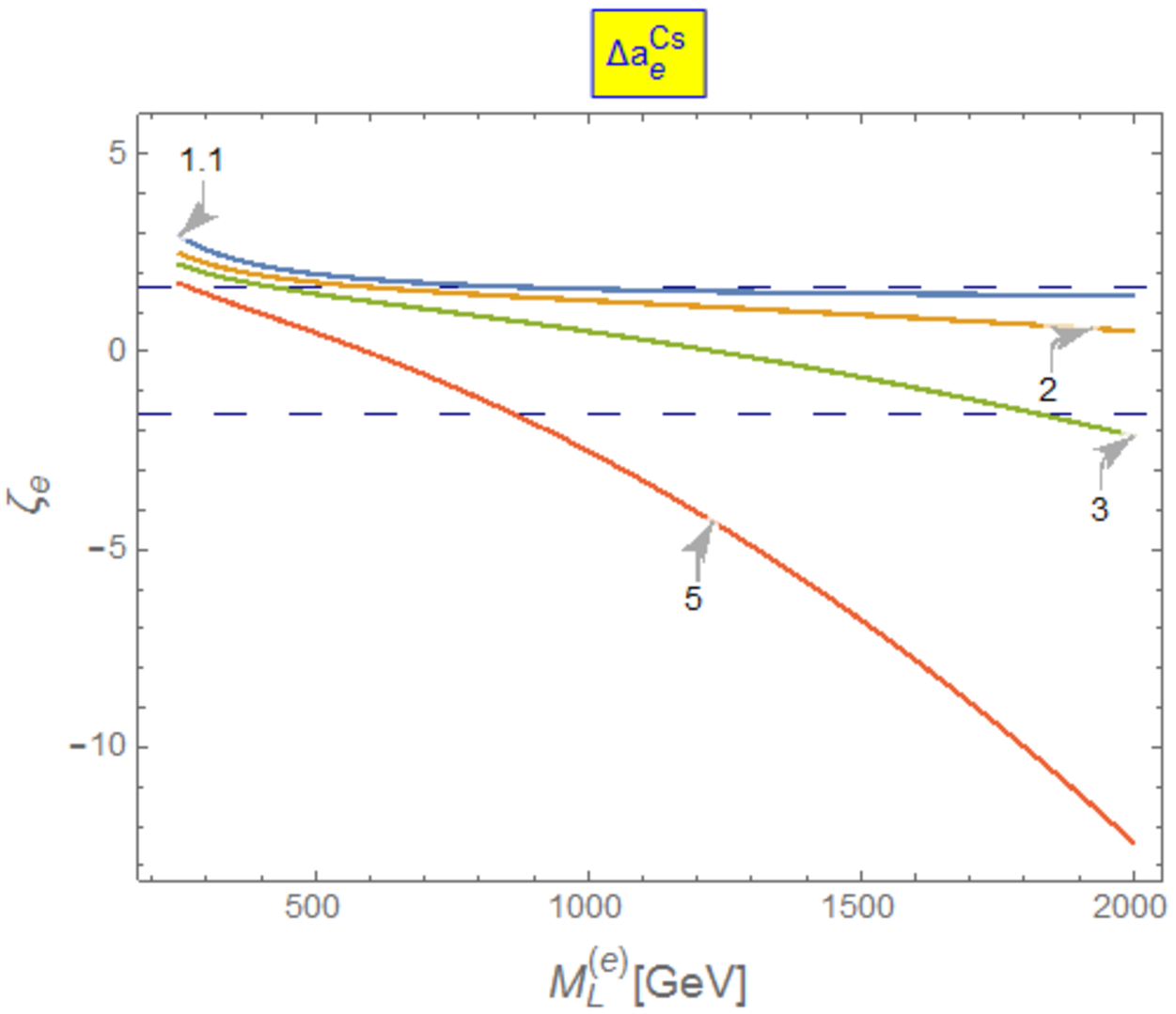} \hspace{0.2cm}
\includegraphics[height=0.22\textheight]{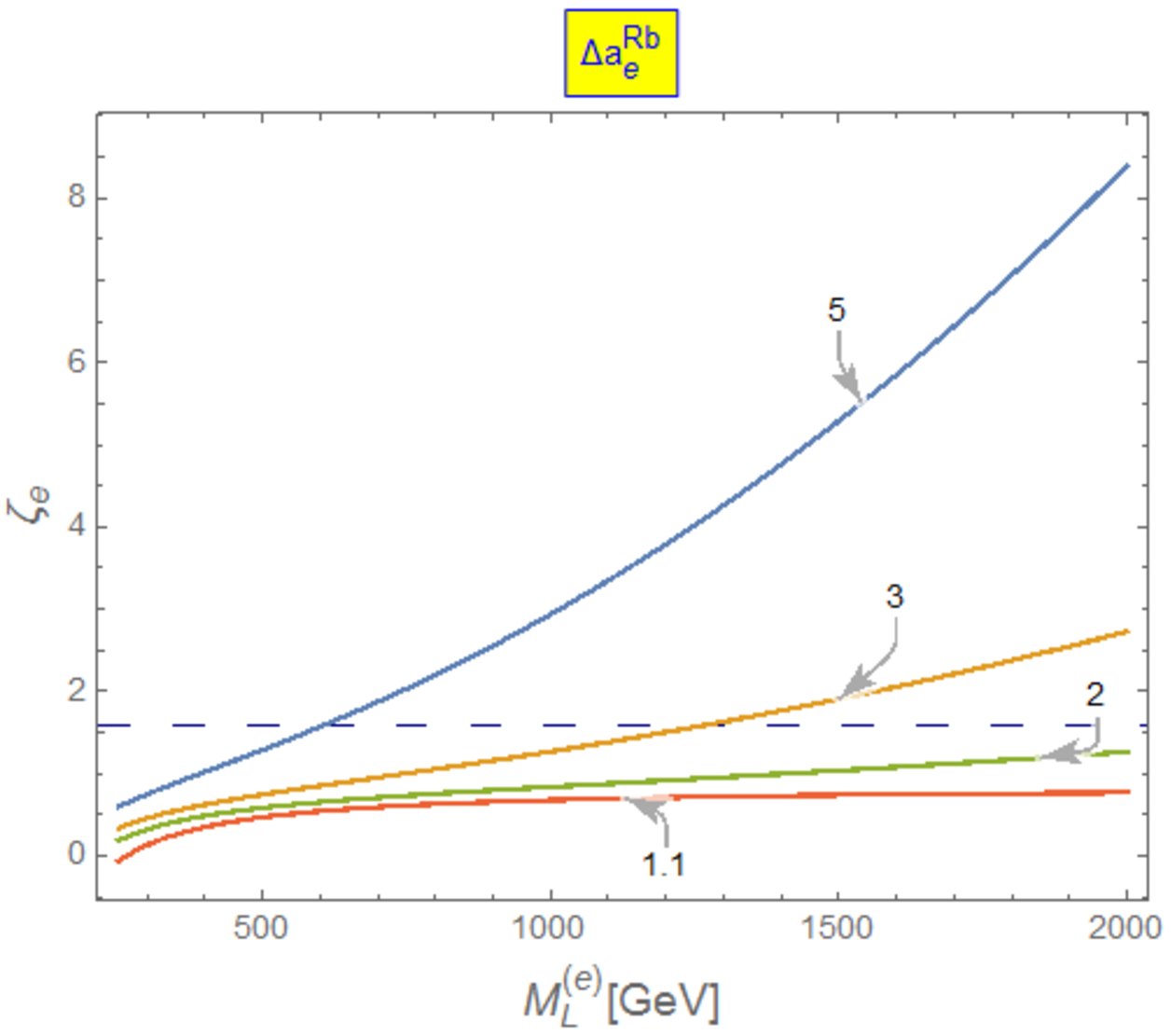}
\caption{ The normalized electron-Higgs Yukawa $\zeta_\mu$ vs $M_L^{(e)}$ for $M_H^{(e)}/M_L^{(e)}=\{1.1,2,3,5\}$. Here we set $M_N=1\mtev$, $\sin(2\alpha^{(e)})=0.3$, $q_h^2= (125 \mgev)^2$, and $\lambda_{H3}=\tilde{\lambda}_{H3}=\lambda_{H6}= 1.0$. The horizontal dashed lines indicate the future experimental sensitivity, $|\zeta_e|<1.6$ at FCC-ee\cite{dEnterria:2021xij}. }
\label{fig:Zeta_E}
\end{figure}

\begin{figure}[htb]
\centering
\includegraphics[height=0.22\textheight]{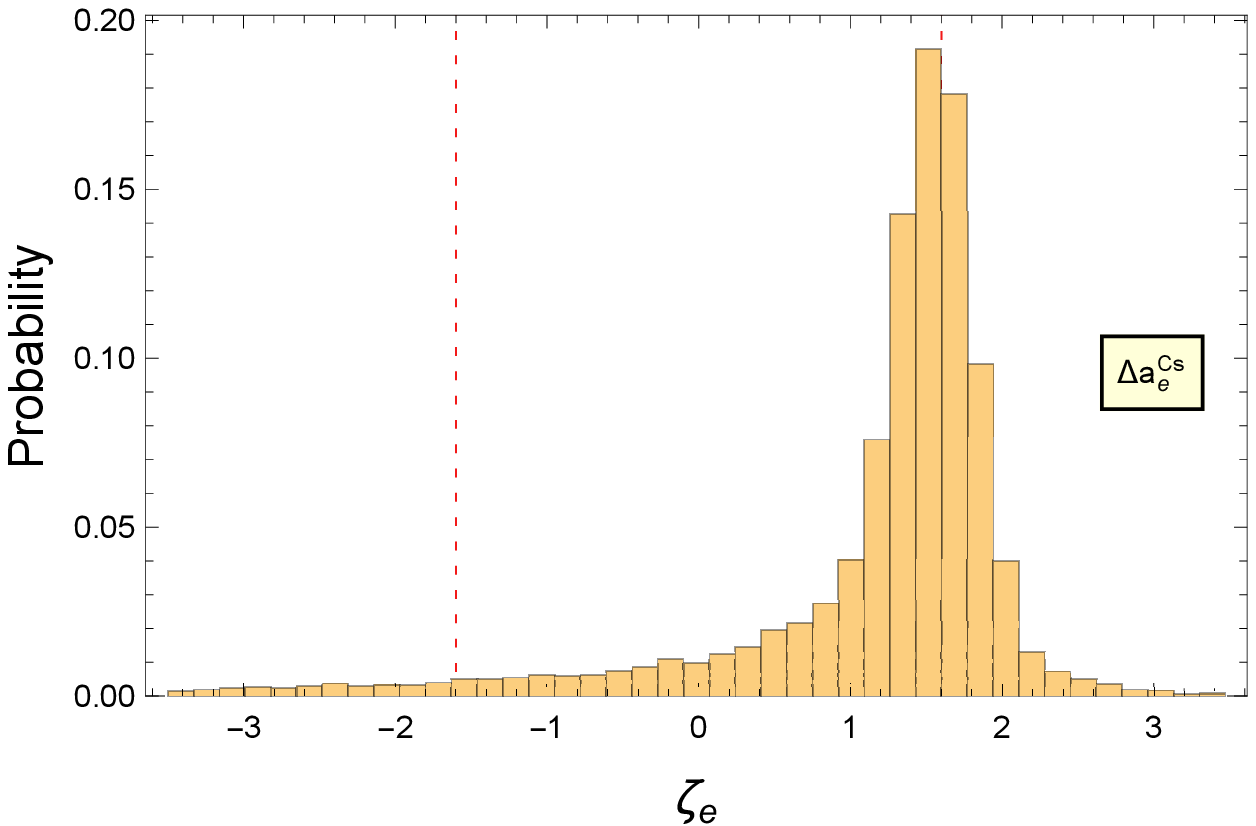}\hspace{0.3cm}
\includegraphics[height=0.22\textheight]{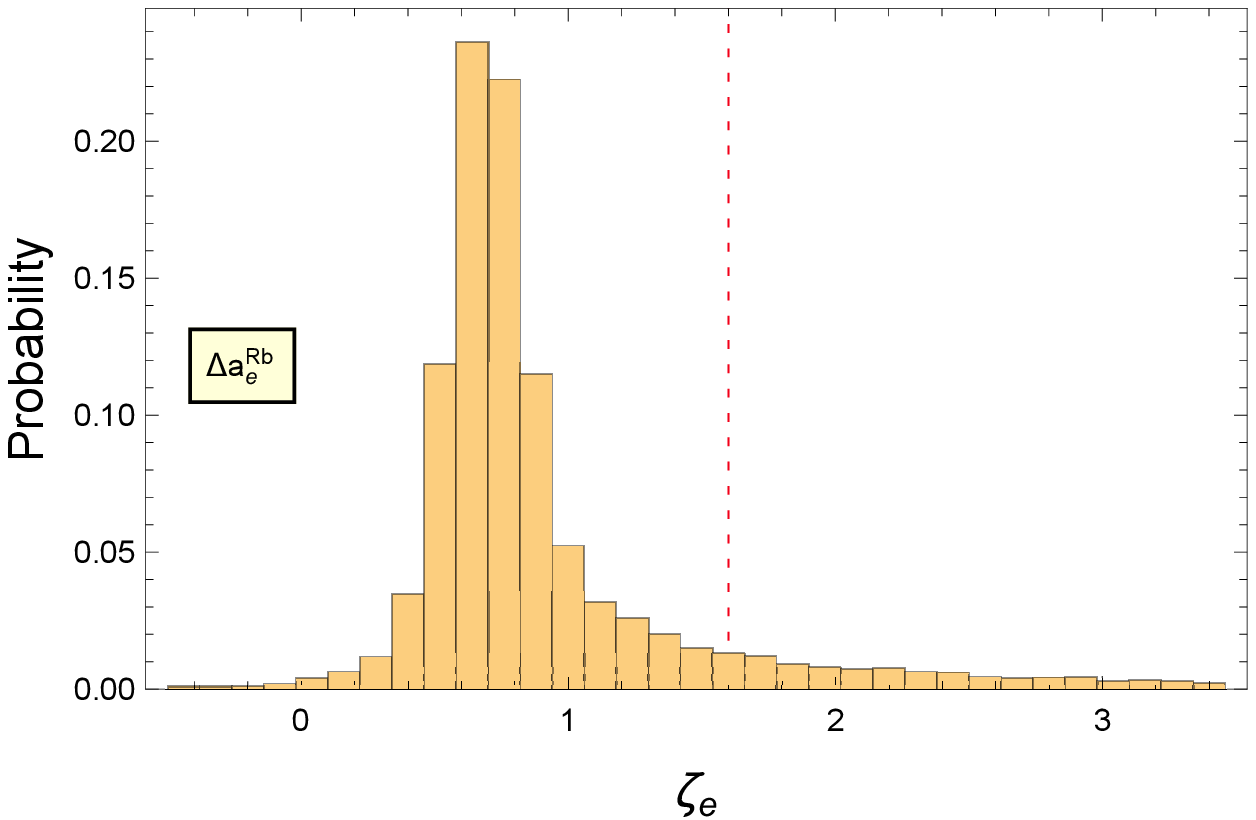}
\caption{ The histogram of the normalized electron-Higgs Yukawa $\zeta_e$.  The vertical dashed lines indicate the future experimental sensitivity, $|\zeta_e|<1.6$ at FCC-ee\cite{dEnterria:2021xij}.  }
\label{fig:Zeta_E_Hist}
\end{figure}

The loop-induced $h_{SM} \bar{e}e$ vertex yields an effective Yukawa coupling. Together with the tree-level contribution, the resulting
effective electron-Higgs Yukawa coupling  is the sum of two.
We take the ratio to the SM electron-(125$\mgev$ Higgs) Yukawa, $y^e_{SM}=m_e/v_0$, and define the normalized electron-Higgs Yukawa as
\beq
\zeta_e={ y^{(e)}_{eff} \over y^e_{SM}} = 1 + \overline{R}_e
\times \left\{-1 + A_0 \!+\! \left[s_2^2-c_2 \left(\lambda_{H3}\! +\!\tilde{\lambda}_{H3}\!-\!\lambda_{H6}\right)\!B_1\right]\! A_1 +
(\lambda_{H3}\! +\!\tilde{\lambda}_{H3}\!+\!\lambda_{H6})\!B_1 A_2\right\}\,.
\eeq
In the above,  the short-handed notations represent
\beqa
s_2&=& \sin(2\alpha^{(e)})\,,\; c_2= \cos (2\alpha^{(e)} )\,,\;
B_1=\frac{v_0^2}{M_N^2}{1 \over (z_H-z_L)}\,,\nonr\\
A_0&=& {{\cal F}\left(z_L,z_H,z_h \right) \over {\cal F}_0\left(z_L,z_H \right) }\,,\nonr\\
A_1&=& {{\cal F}\left(z_H,z_H,z_h \right)+{\cal F}\left(z_L,z_L,z_h \right)  -2 {\cal F}\left(z_L,z_H,z_h \right)\over 2{\cal F}_0\left(z_L,z_H \right) }\,,\nonr\\
A_2&=&  {{\cal F}\left(z_H,z_H,z_h \right)-{\cal F}\left(z_L,z_L,z_h \right) \over 2{\cal F}_0\left(z_L,z_H \right) }\,,
\eeqa
where $z_H=(M^{(e)}_H/M_N)^2$, $z_L=(M^{(e)}_L/M_N)^2$,  $z_h=(q_h^2/M_N^2)$, and $q_h$ is the 4-momentum carried by the SM Higgs.
Numerically, the $z_h$ contribution is insignificant, see Appendix \ref{sec:loop_fun},
$A_0= 1 +{\cal O}(10^{-3})$, $A_1>0$, $A_2<0$, and the absolute values of $A_{1,2}$ increase as the ratio $z_H/z_L$ gets bigger.
We illustrate  the normalized electron-Higgs Yukawa for one particular set of parameters in Fig.\ref{fig:Zeta_E}. One can see that the electron-Higgs Yukawa can be very different, both in magnitude
and sign, from the SM prediction.
To better explore this model, we perform a numerical scan with  $0.2\mtev< M_L <M_H<10\mtev$, $\sin(2 \alpha^{(e)} ) \in [-.3, +.3]$, and each $\lambda_{H3},\tilde{\lambda}_{H3},\lambda_{H6} \in [0.1, \sqrt{4\pi}]$.
The histogram of the resulting $\zeta_e$
is displayed in Fig.\ref{fig:Zeta_E_Hist}, where the vertical dashed lines indicate the projected sensitivity, $|\zeta_e|<1.6$ at $95\%$ CL, at FCC-ee\cite{dEnterria:2021xij}.  With the projected sensitivity at the FCC-ee,
about $40.6(12.6)\%$ of $\zeta_e$ for $\tri a_e^{Cs(Rb)}$ can be detected.
If adopting $\tri a_e^{Cs}$,  $\zeta_e$ spans from $-33.1$ to $5.4$ and peaks in the range $[1.0,2.0](\sim 70\%)$.
The probability for $\{\zeta_e<-5\,,\; \zeta_e <-3\,,\; \zeta_e>3 \}$ are  $\{ 2.3\%, 4.1\%, 0.45\%\}$, respectively.
On the other hand, if $\tri a_e^{Rb}$ is adopted, $\zeta_e$ spans from $-1.4$ to $19.8$ and peaks around $[0.5,1.0](\sim 67\%)$.
The chances for  $\{\zeta_e>5, \zeta_e >3, \zeta_e<-1 \}$ are  $\{ 1.6\%, 4.7\%, 2\times 10^{-4}\}$, respectively.
From the scan, we see that the abnormal electron-Higgs coupling is possible to be tested in the near future.

\begin{figure}[htb]
\centering
\includegraphics[height=0.25\textheight]{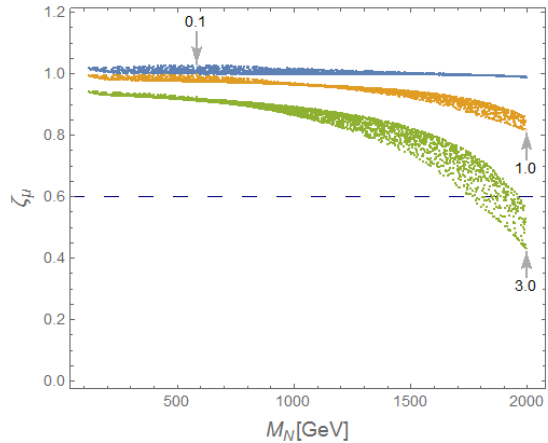}
\caption{ The normalized muon-Higgs Yukawa $\zeta_\mu$ vs $M_N$. Here we set $\tri a_\mu= 25.1\times 10^{-10}$, $\sin(2\alpha^{(\mu)})=0.1$, and $\lambda_{H5}=\tilde{\lambda}_{H5}=\lambda_{H8}=\{ 0.1, 1.0, 3.0\}$. The horizontal dashed line is the current experimental lower limit on $\zeta_\mu>0.6$\cite{ATLAS:2020fzp,CMS:2020xwi}.}
\label{fig:Zeta_Mu}
\end{figure}

\begin{figure}[htb]
\centering
\includegraphics[width=0.5\textwidth]{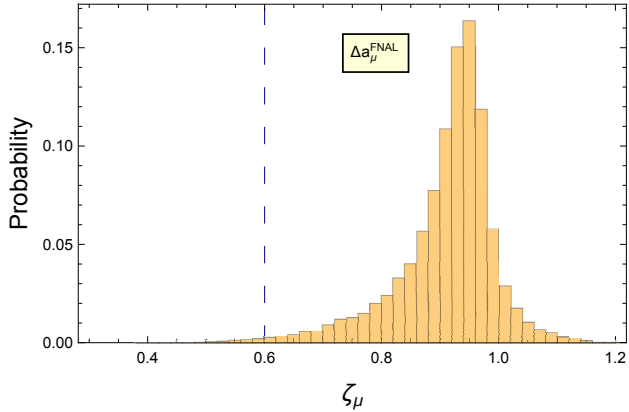}
\caption{ The histogram of the normalized muon-Higgs Yukawa $\zeta_\mu$. The vertical dashed line indicates the current experimental lower limit on $\zeta_\mu>0.6$ \cite{ATLAS:2020fzp,CMS:2020xwi}.  }
\label{fig:Zeta_Mu_Hist}
\end{figure}

On the other hand, muon receives only the radiative mass correction such that $\overline{R}_\mu=1$.
In Fig.\ref{fig:Zeta_Mu}, we display the normalized
muon-Higgs Yukawa for a particular parameter set  $\sin(2\alpha^{(\mu)})=0.1$ and $\lambda_{H5}=\tilde{\lambda}_{H5}=\lambda_{H8}=\lambda_\mu$.
The effective muon-Higgs Yukawa is close to the SM one when $\lambda_\mu=0.1$. If one adopts a larger
coupling $\lambda_\mu$,  $\zeta_\mu<1$, it could be as small as $\sim 0.4$ when $\lambda_\mu=3$ and $M_N= 2\mtev$.
It usually stays within the current $2\sigma$ constraint,  $0.6 \lesssim \zeta_\mu \lesssim 1.5 $\cite{ATLAS:2020fzp,CMS:2020xwi}.
We also perform a numerical scan over the ranges: $M_N \in[0.2,2.0]\mtev$, $\sin(2\alpha^{\mu})\in [-0.3, 0.3]$, and each  $\lambda_{H5},\tilde{\lambda}_{H5},\lambda_{H8} \in[0.1, \sqrt{4\pi}]$.
The histogram of the resulting $\zeta_\mu$ is shown in Fig.\ref{fig:Zeta_Mu_Hist}. The normalized muon Yukawa spans from $0.38$ to $1.21$ and mostly peaks in the range of $[0.8,1.0](\sim82.0\%)$, with the chance of $\sim 7.5\% ( 0.66\%)$, for $\zeta_\mu$ being greater than  $1.0$ (or smaller than the current lower bound $0.6$).
See \cite{deBlas:2019rxi} for the future updates of $\zeta_\mu$ at HL-LHC and $e^+e^-$ colliders.
With the ultima projected precision of $\sim 0.4\%$ at the future FCC\cite{FCC:2018byv}, the entire range of predicted muon-Yukawa
can be covered and probed.

\section{Dark matter and neutrino masses}
\label{sec:DM_Mnu}
The current model does not have a dark matter ( DM ) candidate. However, with the gauged $U(1)_\ell$ symmetry, DM candidate can be easily
included by extending the particle content. One can introduce a pair of vector fermion $D$ which
only interacts with the gauge boson $X$ by adjusting its $U(1)_\ell$ charge, $Q_D$\footnote{ There are infinite possible $Q_D$'s that forbid all Yukawa couplings between $D$ and other fields.  $Q_D=10$ and $Q_D=1/2$  are two concrete examples.  }.
Then $D$ can be assigned with a dark parity without upsetting any gauge symmetry.
This dark parity remains even after the SSB of $U(1)_\ell$ and SM electroweak,  making $D$ a DM candidate.
The annihilation cross section of $D \bar{D} \ra X^*\ra f\bar{f}$ can be calculated to be
\beq
\left\langle \sigma v\right\rangle_{D \bar{D} \ra f\bar{f}}
\simeq Q_D^2 g_l^4{  (G_L^f)^2+(G_R^f)^2 \over \sqrt{2}\pi M_X^2} {\beta_D\over (1-4\beta_D)^2}\,,
\eeq
where $\beta_D=(M_D/M_X)^2$. Since the third generation lepton is $U(1)_l$ neutral, the final states can be $l^+ l^-$ or $\bar{\nu}_l \nu_l$ and $l=e,\mu$.
From $\Omega_{DM}h^2 =0.120\pm 0.001$\cite{Planck:2018vyg} and $\Omega_{DM} h^2 \simeq  4.8 \times 10^{-10} (\mgev)^{-2} / \langle \sigma v\rangle$, we obtain
\beq
\left(\frac{Q_D}{10}\right)^2\left(\frac{g_l}{e}\right)^4 \frac{1}{M_X^2} {\beta_D \over (1-4\beta_D)^2}\simeq 1.1 \times 10^{-10} (\mgev)^{-2}\,.
\eeq
For $M_X=1\mtev$, $Q_D=10$, and $g_l=0.1 e$, either $M_D= 0.394 \mtev$ or $M_D= 0.633 \mtev$ can yield the correct DM relic density.
Moreover, since $D$ is leptophilic, it can safely escape the direct search bound.

Finally, we comment on how to generate the active neutrino masses in this model.
The Majorana neutrino mass matrix element ${\cal M}^\nu_{ij}$ can arise from the corresponding Weinberg operator\cite{Weinberg:1979sa} $(L_i H)(L_j H)$. However, the traditional lepton number is conserved in the current model, which  forbids the Weinberg operator. One possible simple extension is introducing an extra $U(1)_\ell$ charge-2 scalar $S_2$ and allowing it to develop VEV\footnote{This also allows the vector fermion $N$ to acquire a Majorana mass and breaks the traditional lepton number, see\cite{Chang:2022eft}. }. Therefor, the $U(1)_l$ charge of every Weinberg operator can be balanced, and the observed neutrino oscillation data can be explained at the price of potential CLFV. In that case, one must carefully consider the stringent CLFV constraints ( see \cite{Chang:2022eft}, for example), and the comprehensive analysis is beyond the scope of this paper.
\section{Conclusion}
We have proposed an anomaly-free gauged lepton number $U(1)_\ell$ symmetry where the observed $\tri a_{e,\mu}$ are explained, and the charged lepton mass hierarchy arises naturally. On top of the SM particle content, this model requires one pair of vector fermions, two scalar doublets, and three scalar singlets, see Table \ref{table:newparticle}, with TeV-ish masses. The flavor-dependent $U(1)_\ell$ charge assignment, see Table \ref{table:SMparticle}, is novel to our best knowledge.
In our model, tau picks up its mass via the SM Yukawa interaction, while  the $U(1)_\ell$ charge assignment forbids the SM Yukawa interactions for electron and muon.
 Both electron and muon acquire a radiatively generated mass from the photon-removed one-loop diagrams
for the observed $\tri a_{e,\mu}$.
Electron receives an extra mass contribution from its coupling to SM Higgs mediated by one of the exotic doublet scalars.
Compared to the SM tau Yukawa coupling, this model does not require extreme model parameters to reproduce the observed $m_{e,\mu}$ and $\tri a_{e,\mu}$.

This model has two nice features: (1) either positive $\tri a_e^{Rb}\simeq 4.8\times 10^{-13}$\cite{Morel:2020dww} or negative $\tri a_e^{Cs}\simeq -8.7\times 10^{-13}$\cite{Parker:2018vye} can be accommodated with $\tri a_\mu^{B\!N\!L-F\!N\!A\!L} \simeq 25.1 \times 10^{-10}$\cite{Abi:2021gix} in this model, and (2) without any ad hoc symmetries or parities introduced, the automatically emerged conventional lepton number and tau-parity ensure that the tau flavor is conserved to all orders in this model. We have also proved that in the $e-\mu$ sector, all the dim-4 and dim-6 CLFV SM operators vanish to all orders. Hence, no CLFV constraint on this model is expected in the foreseeable future.

We have discussed the phenomenology and pointed out two testable signatures of this model: (1) depending on the overall sign of $U(1)_\ell$ charges, we predicted either $A_{FB}^e >A_{FB}^\tau >A_{FB}^\mu$ or $A_{FB}^e <A_{FB}^\tau <A_{FB}^\mu$. This can be tested at the future $e^+e^-$ colliders  before the direct discovery of the $U(1)_\ell$ gauge boson.
(2) The abnormal  electron- and muon-Higgs Yukawa couplings. Our numerical study found that
$ -33 \lesssim y^{(e)}_{eff}/y^e_{SM} \lesssim 20 $ and  $ 0.6 \lesssim y^{(\mu)}_{eff}/y^\mu_{SM} \lesssim 1.2$, which could be probed in the HL-LHC or future $e^+e^-$ colliders.

Finally, this simple model with this specific $U(1)_\ell$ charge assignment can be ruled out if: (1) any tau flavor violation is confirmed or (2) $|\tri a_\tau|>10^{-10}$ is observed.

\section*{Acknowledgments}
This research is supported by MOST 109-2112-M-007-012 and
110-2112-M-007-028 of Taiwan.

\appendix
\section{Scalar sector}
\label{app:scalarSEC}
Here we spell out the most general scalar sector lagrangian for this model.
The relevant lagrangian can be written as
\beq
{\cal L} \supset {\cal L}_1 -V_1 -V_2\,.
\eeq
The kinetic and quadratic terms are collected in ${\cal L}_1$:
\beqa
{\cal L}_1 &=&  (D_\mu H)^\dag (D^\mu H) + (D_\mu H_5)^\dag (D^\mu H_5) +(D_\mu H_3)^\dag (D^\mu H_3) \nonr\\
&+&(D_\mu C_6)^\dag (D^\mu C_6) +(D_\mu C_8)^\dag (D^\mu C_8) +(D_\mu S_3)^\dag (D^\mu S_3)\nonr\\
&+& \mu^2 H^\dag H -M_5^2 H_5^\dag H_5 -M_3^2 H_3^\dag H_3 -M^2_6 | C_6|^2 -M_8^2 |C_8|^2 + \mu_l^2 |S_3|^2\,,
\eeqa
where the covariant derivative is
\beq
D_\mu = \partial_\mu -i g_2 |T_3|\,\vec{\sigma}\cdot \vec{A}_\mu - i g_1 Y B_\mu -i g_l Q_l X_\mu\,.
\eeq
The gauge invariant renormalizable quartic coupling potential is
\beqa
V_1 &=& \lambda_H (H^\dag H)^2 + \lambda_{H5} (H^\dag H)(H_5^\dag H_5) + \lambda_{H3} (H^\dag H)(H_3^\dag H_3)
+\lambda_{H6} (H^\dag H)|C_6|^2  +\lambda_{H8} (H^\dag H)|C_8|^2 \nonr\\
&+&\lambda_{Hl} (H^\dag H)|S_3|^2
+ \lambda_5 (H_5^\dag H_5)^2 + \lambda_{53} (H_5^\dag H_5)(H_3^\dag H_3) +
\lambda_{56} (H_5^\dag H_5)|C_6|^2  +\lambda_{58} (H_5^\dag H_5)|C_8|^2 \nonr\\
&+&\lambda_{5l} (H_5^\dag H_5)|S_3|^2 + \lambda_3 (H_3^\dag H_3)^2 +
\lambda_{36} (H_3^\dag H_3)|C_6|^2  +\lambda_{38} (H_3^\dag H_3)|C_8|^2 +\lambda_{3l} (H_3^\dag H_3)|S_3|^2\nonr\\
&+&  \lambda_6 |C_6|^4
 +\lambda_{68} |C_6|^2 |C_8|^2 +\lambda_{6l}|C_6|^2 |S_3|^2
  + \lambda_8 |C_8|^4 +\lambda_{8l}|C_8|^2 |S_3|^2
  +  \lambda_l |S_3|^4 \nonr\\
&+& \wt{\lambda}_{H5} (H^\dag H_5)(H_5^\dag H)+ \wt{\lambda}_{H3} (H^\dag H_3)(H_3^\dag H)
+ \wt{\lambda}_{53} (H_5^\dag H_3)(H_3^\dag H_5)\nonr\\
&+&  \bar{\lambda}_{H5} (H^\dag \wt{H}_5)(\wt{H}_5^\dag H)+ \bar{\lambda}_{H3} (H^\dag \wt{H}_3)(\wt{H}_3^\dag H)
+ \bar{\lambda}_{53} (H_5^\dag \wt{H}_3)(\wt{H}_3^\dag H_5)\,.
\eeqa
Also, we have
\beq
V_2= \mu_{H3} H^\dag \wt{H}_3 S^*_3 +\kappa_3 H^\dag H_3 S_3^* C_6 +  \kappa_5 H^\dag H_5 S_3 C_8  +H.c. 
\eeq

From the above, the mass matrix for the charged scalars after SSB can be read
\beq
{\cal L} \supset -(H_3^+,C_6^+) {\cal M}_{36}\left( \begin{array}{c} H_3^- \\ C_6^-\\ \end{array}\right)
-(H_5^+,C_8^+) {\cal M}_{58}\left( \begin{array}{c} H_5^- \\ C_8^-\\ \end{array}\right)\,,
\eeq
where
\beq
{\cal M}_{36}= \left( \begin{array}{cc} M_3^2+\frac{\lambda_{H3}+\wt{\lambda}_{H3}}{2} v_0^2 +\frac{\lambda_{3l}}{2} v_3^2 & \frac{1}{2}\kappa_3 v_0 v_3 \\
\frac{1}{2}\kappa_3 v_0 v_3 & M_6^2+\frac{\lambda_{H6}}{2} v_0^2 +\frac{\lambda_{6l}}{2} v_3^2 \\ \end{array}\right)\,,
\eeq
and a similar form for ${\cal M}_{58}$.
Also, the couplings between the charged scalars  and the neutral component of the SM Higgs doublet $h$ is
\beq
{\cal L} \supset - h (H_3^+,C_6^+)\Lambda_{36}\left( \begin{array}{c} H_3^- \\ C_6^-\\ \end{array}\right)\,,\;\;
\Lambda_{36}= \left( \begin{array}{cc} (\lambda_{H3}+\wt{\lambda}_{H3}) v_0 & \frac{1}{2}\kappa_3 v_3 \\
\frac{1}{2}\kappa_3 v_3 & \lambda_{H6}v_0  \\ \end{array}\right)\,.
\eeq

\section{1-loop function}
\label{sec:loop_fun}
When evaluating the Feynman diagrams displayed in Fig.\ref{fig:g-2_mass}, one encounters the following integral
\beq
{\cal F}(a,b,c )=\int_0^1 dx \int^{1-x}_0 dy {1\over 1-x-y +a x +b y -x y c}\,,
\eeq
where $a,b,c \geq 0$.
As long as $c<4\sqrt{a b}$, ${\cal F}(a,b,c )$ is always positive.
For $c\ll a, b$, the loop integral can be expanded as
\beq
{\cal F}(a,b,c )= {\cal F}_0(a,b)+ c\, {\cal F}_1(a,b)  +{\cal O}(c^2)\,,
\eeq
where
\beqa
{\cal F}_0(a,b)&=&\frac{1}{a-b}\left( \frac{ a\ln a}{a-1} - \frac{ b\ln b}{b-1} \right)\,,\nonr\\
{\cal F}_1(a,b)&=& \int_0^1 dx \int^{1-x}_0 dy {x y \over (1-x-y +a x +b y)^2 }\,.
\eeqa
For the parameter space we are interested in, $0.2\,\mtev<M_N<2\,\mtev$ and $0.5\,\mtev< M_{L_l}<M_{H_l}<\,10\mtev$,  $10^{-5}< {\cal F}_1/{\cal F}_0<0.5$.
Moreover,
\beq
{\cal F}_0(a,b)\ra {a-1 -\ln a \over (a-1)^2}\,,\;\;  \mbox{when } a \ra b\,.
\eeq

\section{Forward-backward asymmetry}
\label{sec:appendAFB}
The tree level forward-backward asymmetry of $e^+e^- \ra f \bar{f}$
due to the interference among photon, $Z$, and $X$
can be easily calculated and summarized as
\beqa
A^f_{FB}(s)&\equiv& { \sigma^F(e^+e^- \ra f \bar{f}) - \sigma^B(e^+e^- \ra f \bar{f})\over \sigma^F(e^+e^- \ra f \bar{f}) + \sigma^B(e^+e^- \ra f \bar{f})}\nonr\\
&=&\frac{3}{4}{\sum_{a,b=\gamma,Z,X} K_{ab}(s)(L_{ab}^e- R_{ab}^e)(L_{ab}^f-R_{ab}^f) \over \sum_{a,b=\gamma,Z,X} K_{ab}(s)(L_{ab}^e+R_{ab}^e)(L_{ab}^f+R_{ab}^f) }\,.
\eeqa
The interfering terms are given by
\beq
K_{ab}(s) =(2-\delta_{ab})\Re\left[ \left( 1-\frac{M^2_a}{s} +i \frac{M_a \Gamma_a}{s}\right)^{-1}
\left( 1-\frac{M^2_b}{s} -i \frac{M_b \Gamma_b}{s}\right)^{-1} \right]\,,
\eeq
where $s$ is the CM energy squared, and the Kronecker delta function takes care of the proper factors.
The short handed notations are defined as
\beq
L^f_{ab}=G^f_{aL} G^f_{b L}\,,\; R^f_{ab}=G^f_{aR} G^f_{b R}\,,
\eeq
where $G_{L/R}$ is the gauge couplings normalized to $e$. Namely,
 $G^f_{\gamma, L/R}=Q^f$, $G^f_{Z, L/R} =(T_3-Q)^f_{L/R}/(c_W s_W)$, and $G^f_{X, L/R}= (g_l/e) Q^f_{l L/R}$.

\bibliography{gaugedL_Ref}
\end{document}